\documentclass[a4paper]{jpconf}
\usepackage{graphicx}

\usepackage{epstopdf}
\usepackage{xcolor}

\usepackage{dcolumn}
\usepackage{bm}
\usepackage{amssymb}
\usepackage{xspace}

\usepackage{hyperref}
\usepackage{amsmath}
\usepackage{lineno}
\usepackage{float}

\newcommand{\pp}{\ensuremath{\rm pp}\xspace}

\newcommand{\raa}{\ensuremath{R_{\rm AA}}\xspace}

\newcommand{\pbpb}{Pb--Pb\xspace}
\newcommand{\auau}{Au--Au\xspace}

\newcommand{\ppb}{p--Pb\xspace}

\newcommand{\snnt}[1]{\ensuremath{\sqrt{s_{\rm NN}} = #1 \text{\,TeV}}\xspace}

\newcommand{\sppt}[1]{\ensuremath{\sqrt{s} = #1 \text{\,TeV}}\xspace}

\newcommand{\pT}{\ensuremath{p_{\rm{T}}}} 

\begin{document}
\title{Review of recent results on heavy-ion physics and astroparticle physics in ALICE at the LHC}

\author{H\'ector Bello and Arturo Fern\'andez}

\address{Facultad de Ciencias F\'isico Matem\'aticas BUAP, 1152, Puebla, M\'exico.}

\author{Antonio Ortiz}

\address{Instituto de Ciencias Nucleares, UNAM, Ciudad de M\'exico, 04510, M\'exico}

\ead{antonio.ortiz@nucleares.unam.mx}

\begin{abstract}

In this work we present a summary of the most relevant results on heavy-ion and astroparticle physics in ALICE. The summary includes a brief overview of the current status on the characterization of the hot and dense QCD medium created in the heavy-ion collisions produced at the LHC, as well as the intriguing finding of collective-like phenomena in small collision systems. 

\end{abstract}

\section{Introduction}
ALICE (A Large Ion Collider Experiment~\cite{0954-3899-32-10-001}) is a dedicated heavy-ion detector to exploit the unique physics potential of nucleus-nucleus interactions at the Large Hadron Collider (LHC) energies. The main goal of the detector is to study the physics of strongly interacting matter at extreme energy densities, where we have evidence that a new phase of matter, the quark-gluon plasma (QGP), is formed~\cite{Adams:2005dq,Back:2004je,Arsene:2004fa,Adcox:2004mh}. Although several important measurements have been carried out by the LHC high energy physics experiments (CMS, ATLAS and LHCb), the present work focuses on results from ALICE and it is based on a recent review on heavy-ion physics at the LHC~\cite{Bala:2016hlf}. In this paper, we discuss the discovery of QGP-like phenomena in small collision systems, proton-proton (\pp) and proton-led (\ppb) collisions. And also, we briefly discuss other contributions of ALICE in astroparticle physics.


\section{The ALICE apparatus}
\label{alice}

Particle identification (PID) is an important tool to study the hot and dense matter created in heavy-ion collisions. Therefore, ALICE~\cite{Abelev:2014ffa} is an experiment specialized in PID from low ($\approx 0.15$  GeV/$c$) up to high (20 GeV/$c$) transverse momenta. The central barrel of ALICE is placed inside a large solenoidal magnet which provides a magnetic field of 0.5 T. It is dedicated to detect hadrons, electrons, and photons produced at mid-pseudorapidity, $|\eta|$$<$0.8. It comprises an Inner Tracking System (ITS) of high-resolution silicon detectors, a cylindrical Time-Projection Chamber (TPC), and particle identification arrays of Transition-Radiation Detectors (TRD) and of Time-Of-Flight (TOF) counters. Additional central subsystems, not-covering full azimuth, are a ring-imaging Cherenkov detector for High-Momentum Particle IDentification (HMPID), and two electromagnetic calorimeters: a high-resolution PHOton Spectrometer (PHOS) and a larger-acceptance ElectroMagnetic Calorimeter (EMCal). The muon arm detects muons emitted within  $2.5< \eta <4$ and consists of a complex arrangement of absorbers, a dipole magnet, five pairs of tracking chambers, and two trigger stations. Several smaller detectors (VZERO, TZERO, FMD, ZDC, and PMD) for triggering, multiplicity measurements and centrality determination are installed in the forward region. On the  three upper faces of the octagonal yoke of the solenoid, a dedicated cosmic ray detector (ACORDE~\cite{FernandezTellez:2007yxa}) is placed, which is made of 60 scintillator modules, covering 10\% of solenoid upper faces. This subsystem  produces a cosmic ray trigger signal  when a single module fires (single muon cosmic ray event) or when more than two modules are activated by a cosmic ray shower.



\section{Characterization of QGP}

\subsection{Soft probes}

To learn about the early stage of the system, low transverse momentum ($p_{\rm T}$ $<2.2$\,GeV/$c$) direct photons are studied. A temperature $T$~=~297$\pm$12$^{\rm stat}$$\pm$41$^{\rm syst}$ MeV has been measured for the 0-20\% \pbpb collisions at $\sqrt{s_{\rm NN}}=$2.76\,TeV~\cite{Adam:2015lda}. Hence, the system at the LHC is hotter than that produced at RHIC, where an early temperature of 239$\pm$25$^{\rm stat}$$\pm$7$^{\rm syst}$ MeV was measured for the same centrality class in \auau collisions~\cite{Adare:2014fwh}. It is worth noticing that such temperatures are already above the one predicted to achieve the QCD phase transition~\cite{Ayala:2014jla}.  In addition, the average multiplicity per number of participant is twice that measured at RHIC~\cite{Aamodt:2010cz}.

The system expands and cools down, then, the inelastic interactions cease and the yields of particles reach fixed values. This is the stage of the so-called chemical freeze-out which is studied using the yields of identified hadrons.  Within 20\% particle ratios, e.g., the proton yield normalized to that of pions, are described by thermal models with a common chemical freeze-out temperature of $T_{\rm ch}\approx$ 156 MeV~\cite{Floris:2014pta}. However, large deviations are observed for protons and $\rm K^{*0}$; for the latter, this is not a surprise since its mean lifetime is smaller than that of the fireball ($\approx$10 fm/$c$)~\cite{Aamodt:2011mr}, and therefore the resonance yield may deviate from the expected values  due to hadronic processes like re-scattering and regeneration~\cite{Abelev:2014uua}. 

\begin{figure*}[t!]
\begin{center}
\includegraphics[keepaspectratio, width=0.5\columnwidth]{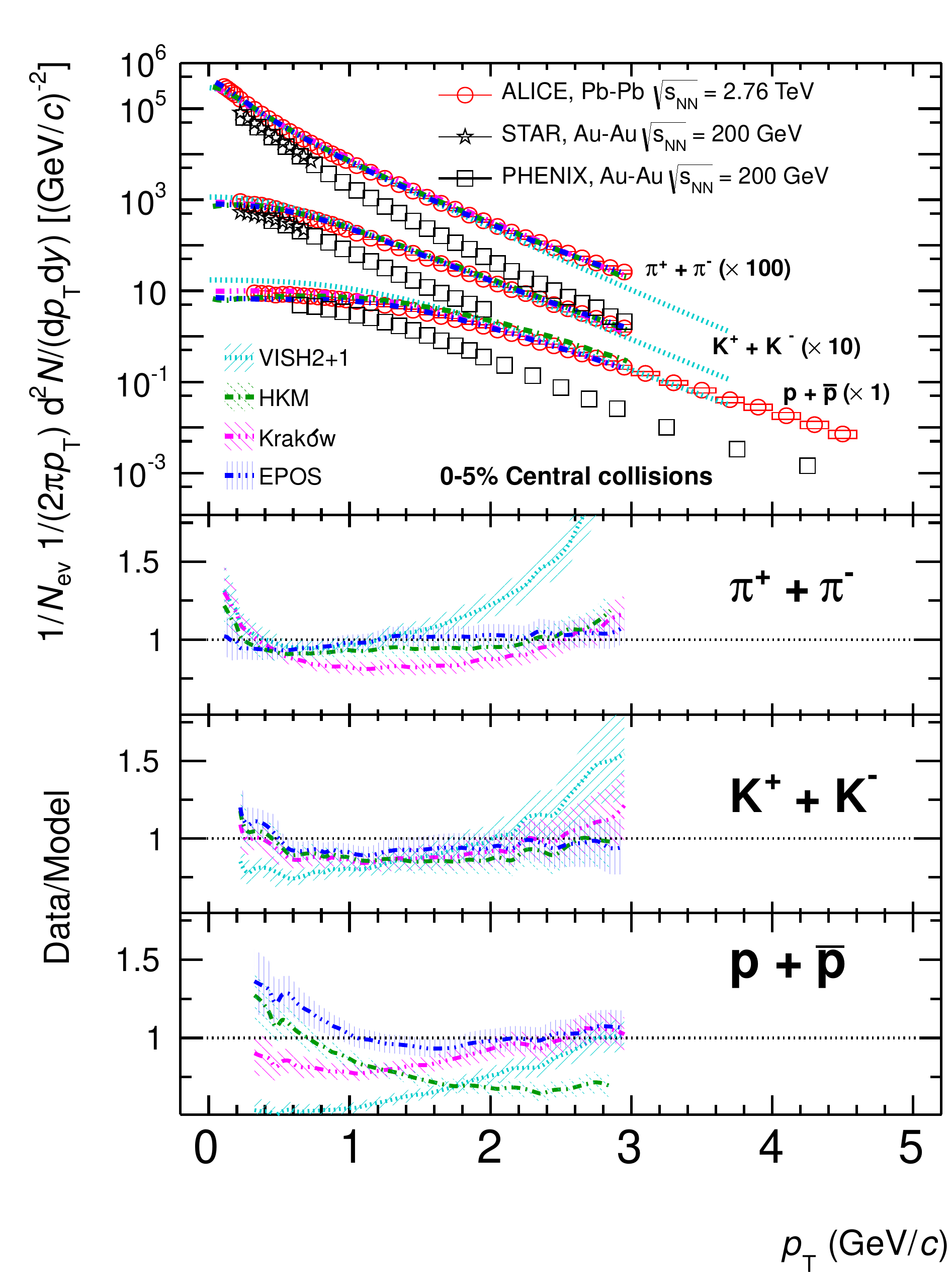}
\includegraphics[keepaspectratio, width=0.5\columnwidth]{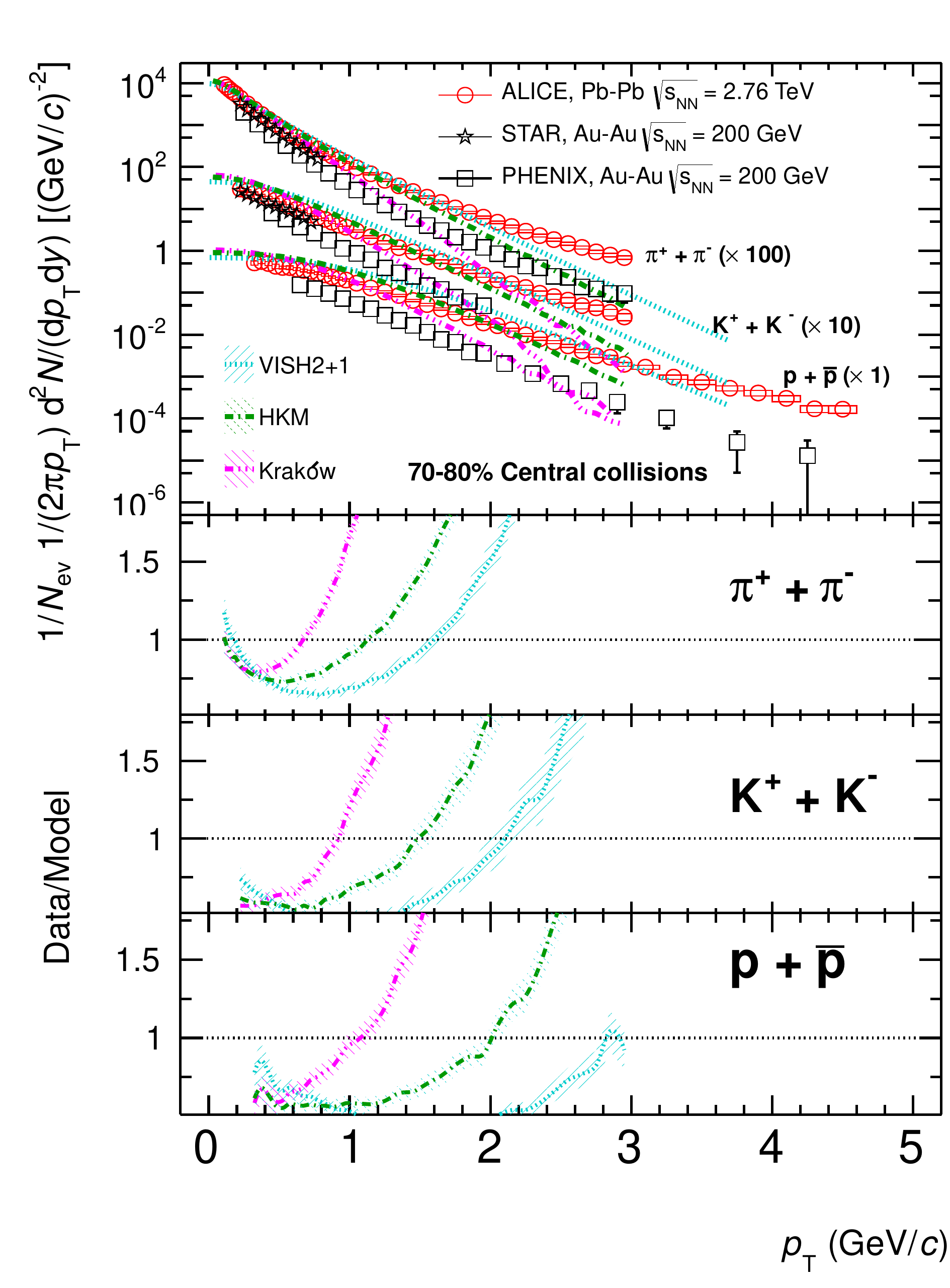}
\caption{\label{fig:pbpb:1} (Color online) Transverse momentum spectra of charged pions, kaons, and (anti)protons measured in central (left) and peripheral (right) \pbpb collisions at $\sqrt{s_{\rm NN}}=2.76$\,TeV. The systematic and statistical uncertainties are plotted as color boxes and vertical error bars, respectively.}
\end{center}
\end{figure*}

On the other hand, the measurement of the spatial extent at decoupling time is obtained using intensity interferometry, a technique which exploits the Bose-Einstein enhancement of identical bosons emitted close by in phase space. This approach is known as Hanbury Brown-Twiss analysis (HBT)~\cite{HanburyBrown:1956bqd}. Using identical charged pions ALICE has measured a homogeneity volume of $\approx$ 300\,fm$^{3}$ (two times that reported at RHIC) and a decoupling time of $\approx$ 10\,fm/$c$~\cite{Aamodt:2011mr}.

The transverse momentum distributions of identified hadrons contain valuable information about the collective expansion of the system ($p_{\rm T}$~$\lesssim$~2\,GeV/$c$), the presence of new hadronization mechanisms like quark recombination (2~$\lesssim$~$p_{\rm T}$~$\lesssim$~8\,GeV/$c$)~\cite{Fries:2008hs} and, at larger transverse momenta, the possible modification of the fragmentation due to the medium~\cite{Sapeta:2007ad,Bellwied:2010pr}. ALICE has reported the  transverse momentum spectra of charged pions, kaons and (anti)protons as a function of the collision centrality in broad transverse momenta intervals~\cite{Abelev:2013vea,Abelev:2014laa,Adam:2015kca}.

Figure~\ref{fig:pbpb:1} shows that for central \pbpb collisions the low $\pT$  parts ($<$~2\,GeV/$c$) of the spectra are well described by hydrodynamic models (within 20\%), except the low $p_{\rm T}$ ($<$~1\,GeV/$c$) proton yield~\cite{Bozek:2012qs,Karpenko:2012yf,Werner:2012xh,Shen:2011eg}. Models which best describe the data include hadronic rescattering with non-negligible antibaryon annihilation~\cite{Werner:2012xh,Shen:2011eg}. The description of the results by hydrodynamic models is only observed in 0-40\% \pbpb collisions, results for more peripheral collisions disagree with such prediction. This behavior has been recently studied for the average $p_{\rm T}$ in different colliding systems~\cite{Ortiz:2015cma}. 

The analysis of the spectral shapes of the $p_{\rm T}$ distributions can be done using a blast-wave fit~\cite{PhysRevC.48.2462}, which allows the extraction of the parameters related with the temperature at the kinetic freeze-out ($T_{\rm kin}$) and the average transverse expansion velocity ($\langle \beta_{\rm T} \rangle$). At the LHC, the radial flow in the most central collisions is found to be $\approx$ 10\% higher than at RHIC,  while the kinetic freeze-out temperature was found to be close to that measured at the RHIC, $T_{\rm kin}=$~95\,MeV~\cite{Abelev:2013vea}. From the study of the low $p_{\rm T}$ particle production we conclude that at the LHC the created system is larger, hotter and longer-lived than that produced at RHIC.

\begin{figure*}[t!]
\begin{center}
\includegraphics[keepaspectratio, width=0.8\columnwidth]{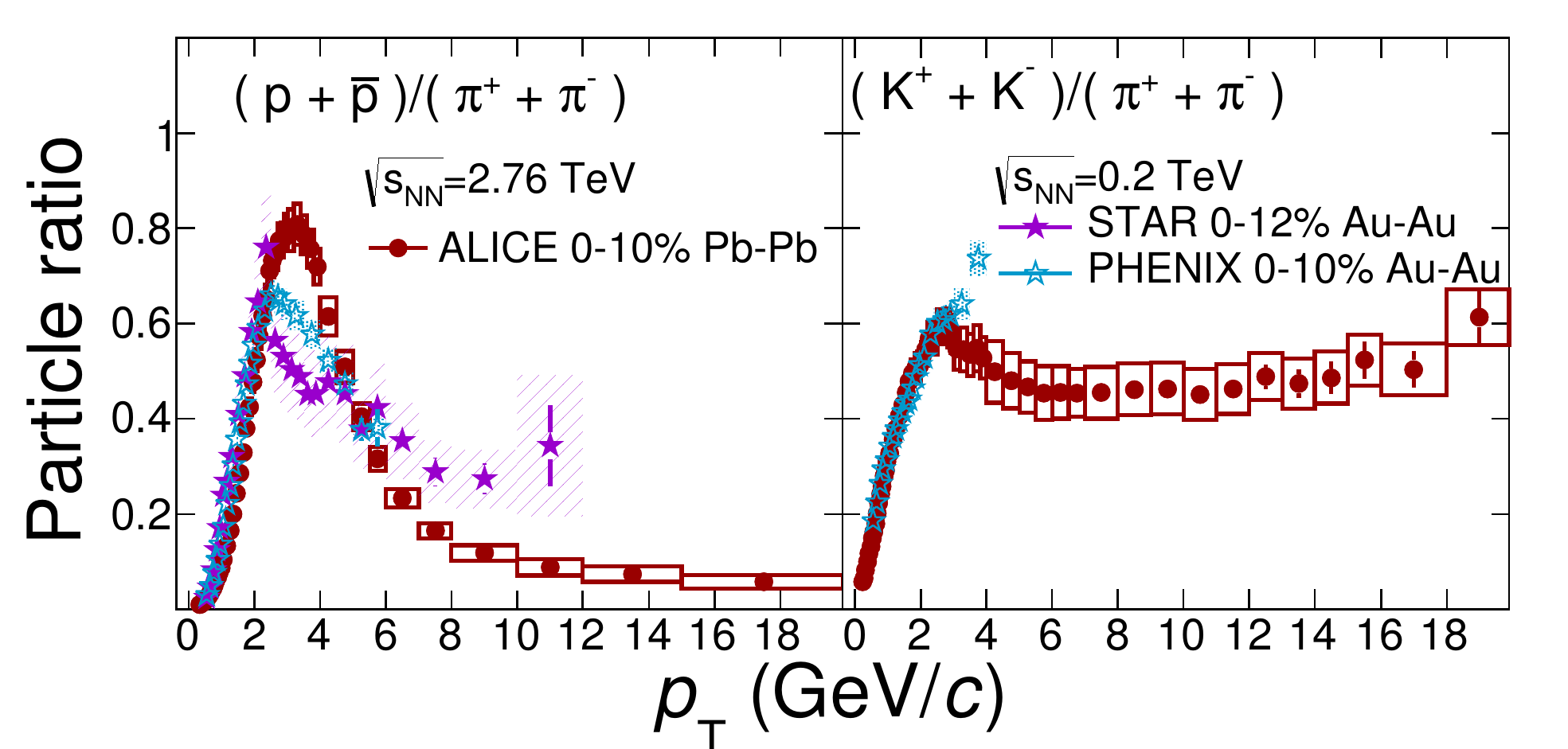}
\caption{\label{fig:pbpb:2} (Color online) ALICE (circles) results from $\sqrt{s_{\rm NN}}=2.76$\,TeV \pbpb collisions compared with STAR
and PHENIX results for $\sqrt{s_{\rm NN}}=200$\,GeV \auau collisions.  Left panel:  the proton-to-pion ratio.  Right panel: the kaon-to-pion ratio.}
\end{center}
\end{figure*}

The intermediate $p_{\rm T}$ is studied with the particle ratios as a function of $p_{\rm T}$~\cite{Adam:2015kca}. For central heavy-ion collisions, Fig.~\ref{fig:pbpb:2} shows a comparison of particle ratios with results from STAR and PHENIX at the RHIC measured in \auau collision at
$\sqrt{s_{\rm NN}}=200$\,GeV. The proton-to-pion peak at the LHC is approximately 20\% larger than at the RHIC, which is consistent with an increased average radial flow velocity. Interestingly, in ALICE 
we also observe a peak in the kaon-to-pion ratio supporting the idea of a strong radial flow.
To check the effect of radial flow the shapes of the $p_{\rm T}$ distributions of $\phi$-meson and protons are compared. The results indicate that for central \pbpb collisions the shapes of the proton-to-pion and $\phi$-to-pion ratios are the same.  On the other hand, for $p_{\rm T}<$~4\,GeV/$c$, the $\phi$-meson yield normalized to that of protons becomes flat going from the most peripheral to the most central \pbpb collisions. This suggests that the mass, and not the number of quark constituents, determines the spectral shape in central \pbpb collisions. This is in a good agreement with the hydrodynamical interpretation. Recently, it has been shown that the spectral shapes, studied with the average $p_{\rm T}$, exhibit a scaling with the hadron mass (number of constituent quarks) only in the 0-40\% (40-90\%) \pbpb collisions~\cite{Ortiz:2015cma}.

Using the scalar product method, the elliptic flow for identified hadrons has been  measured~\cite{Abelev:2014pua} over a broad $p_{\rm T}$ range. Going from central to semi-peripheral \pbpb collisions, the second order Fourier coefficient ($v_{2}$) increases as expected due to the eccentricity increase. For $p_{\rm T}$ below 2\,GeV/$c$ a mass ordering is observed indicating the interplay between elliptic and radial flow. For higher $p_{\rm T}$, the hadron-$v_{2}$ seems to be grouped into baryons and mesons, the exception is the $v_{2}$ of $\phi$-mesons, which for central \pbpb collisions follows that for baryons. This observation indicates that the behavior of $v_{2}$ is driven by the hadron mass and not by the number of constituent quarks. ALICE has also reported the violation of the scaling of $v_{2}$ with the number of constituent quarks, such a observation is also against the scenario with quark recombination/coalescence.

\subsection{Jet studies}

The amount of suppression of inclusive particle production relative to pp collisions can be quantified with the nuclear modification factor, \raa, defined as 
\begin{equation}\label{eq:res:1}
R_{\rm AA}(p_{\rm T}) = \frac{1}{\langle T_{\rm AA} \rangle} \frac{{\rm d^{2}}N^{AA}/{\rm d}y{\rm d}p_{\rm T}}{{\rm d^{2}}\sigma^{pp}/{\rm d}y{\rm d}p_{\rm T}}
\end{equation}
where $N^{\rm AA}$ ($\sigma^{\pp}$) is the yield (cross section) measured in AA (\pp) collisions; and  $\langle T_{\rm AA} \rangle$ is the nuclear overlap function~\cite{Abelev:2013qoq}.

\begin{figure}[H]
\begin{center}
\includegraphics[keepaspectratio, width=0.7\columnwidth]{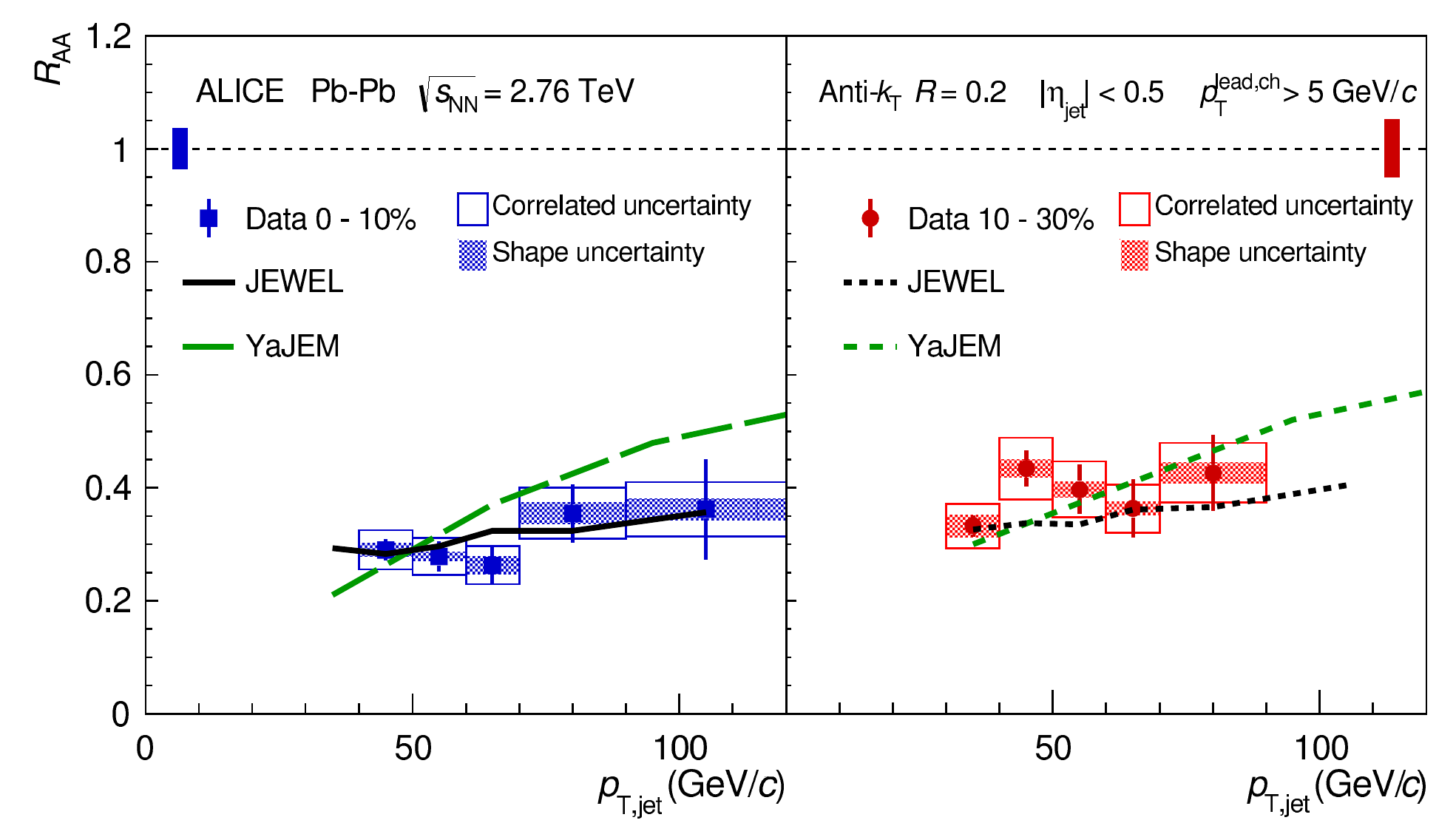}
\caption{\label{fig:fulljetRAA} (Color online) Nuclear modification factor of full jets with $R$~=~0.2 and $p_{\rm T}$ leading bias of 5\,GeV/$c$ in central \pbpb collisions~\cite{Adam:2015ewa}.}
\end{center}
\end{figure}

Figure~\ref{fig:fulljetRAA} shows $R_{\rm AA}$ of full jets with $R$~=~0.2--0.3 in central \pbpb collisions at $\snnt{2.76}$. The data exhibit a strong suppression of the jet production relative to \pp collisions~\cite{Abelev:2013fn}.  Comparing the nuclear modification factor of charged jets to that of charged particles at large $p_{\rm T}$ one can see that the amount of the suppression is similar although the underlying parton $p_{\rm T}$ scale is different for inclusive particles and jets.

In order to study details of path-length dependence of energy loss, ALICE performed studies of elliptic anisotropy of inclusive charged jets~\cite{Adam:2015mda} and semi-inclusive distributions of recoil jets~\cite{Adam:2015doa} which complement and further extend earlier studies of elliptic anisotropies of inclusive high-$p_{\rm T}$  particles and modification of away-side di-hadron correlations~\cite{Aamodt:2011vg}. For collisional energy loss, the path length dependence is expected to be linearly proportional to the length traversed by the parton in medium, while for radiative energy loss, where in addition  interference effects play a role, the dependence can be quadratic. In AdS/CFT class of models  an  even stronger dependence on path length traversed is predicted. 

\begin{figure}[H]
\begin{center}
\includegraphics[keepaspectratio, width=0.45\columnwidth]{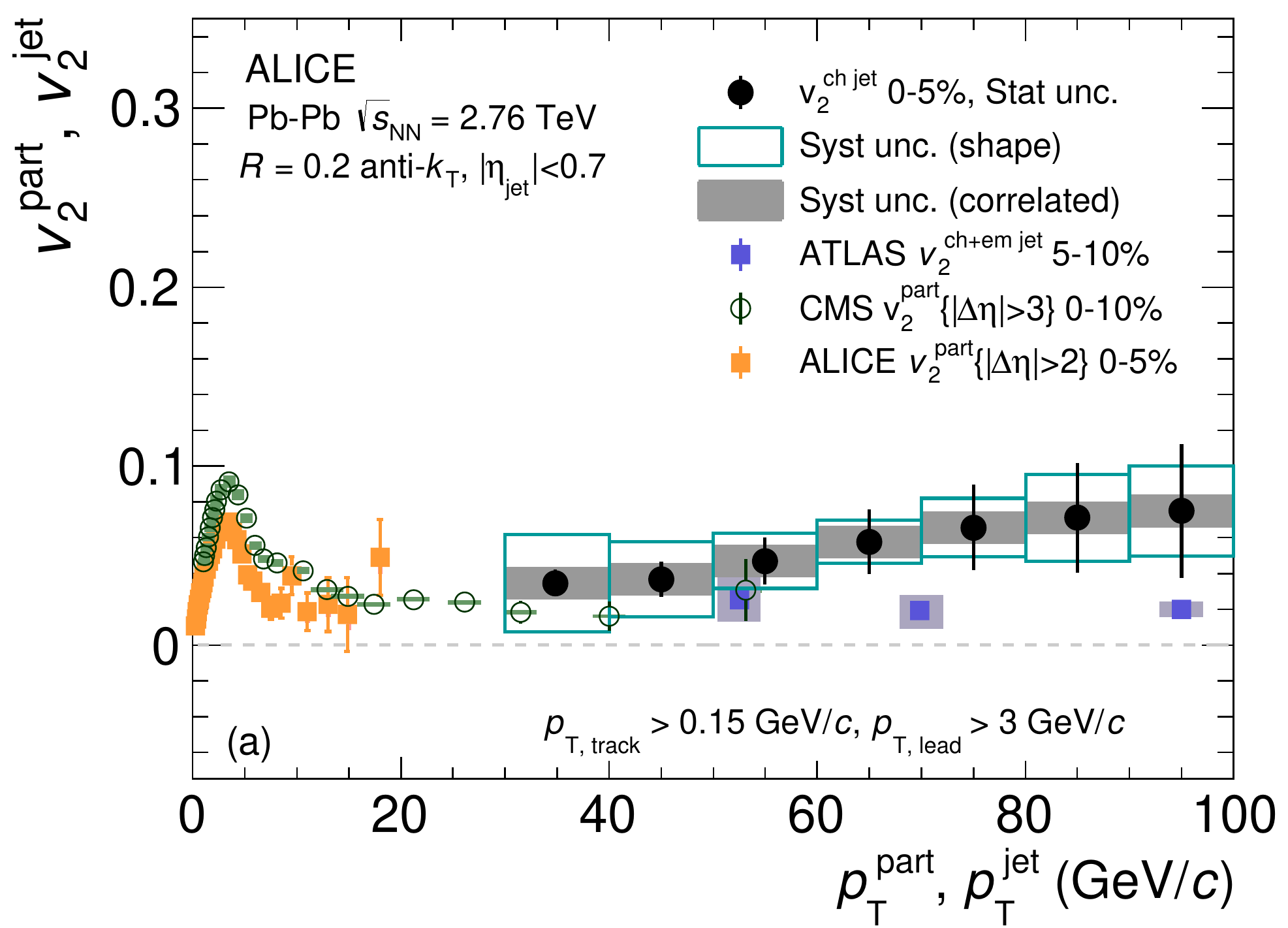}
\includegraphics[keepaspectratio, width=0.45\columnwidth]{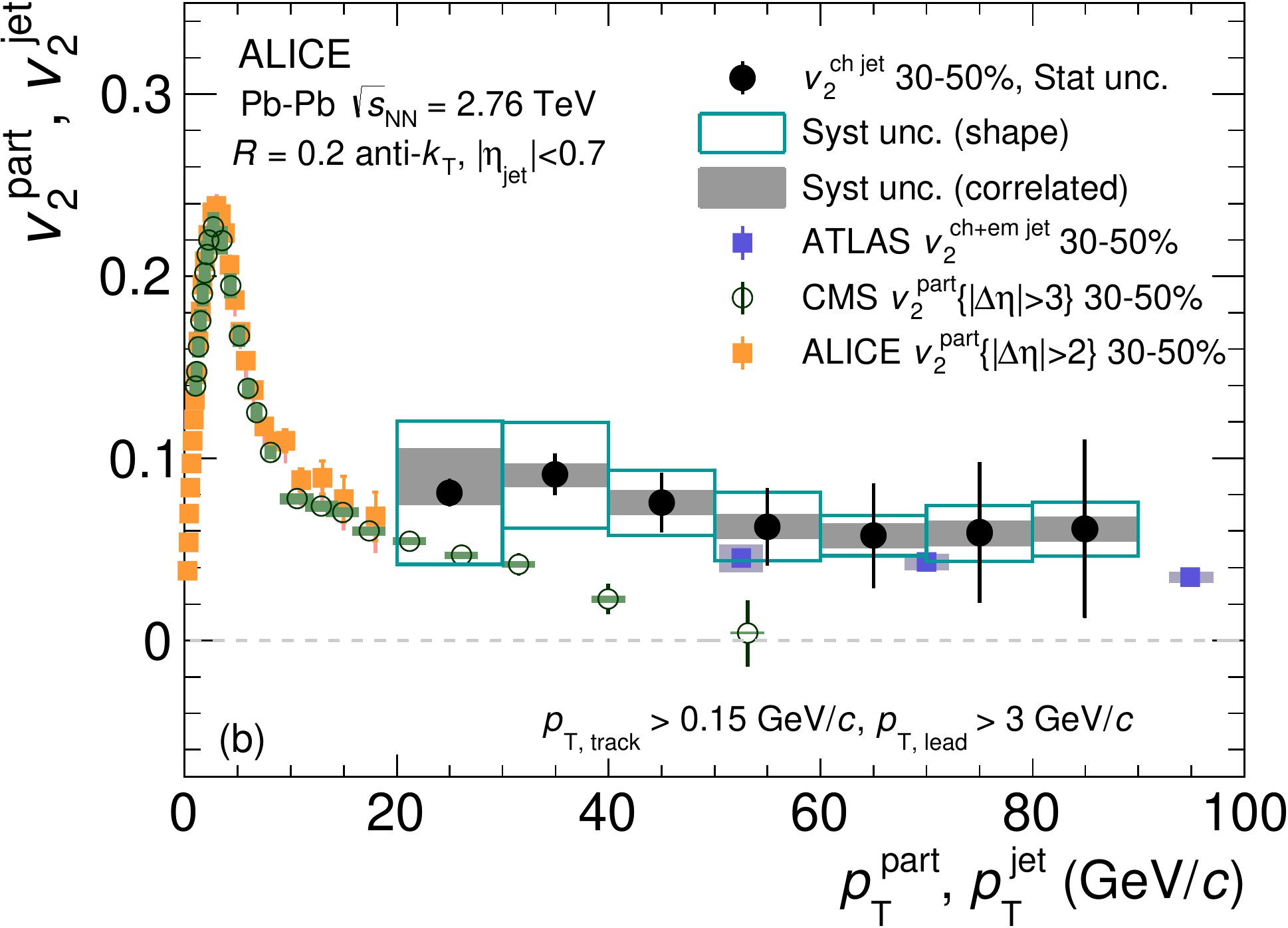}
\caption{\label{fig:v2jetALICE} (Color online)  Elliptic anisotropy ($v_2$) of inclusive charged particles and charged jets with the resolution parameter $R$~=~0.2 in central (left) and semicentral (right) \pbpb\ collisions at $\snnt{2.76}$ measured by ALICE~\cite{Adam:2015mda}. The data are compared with charged particle $v_2$ anisotropies measured by ALICE~\cite{Abelev:2012di} and CMS~\cite{Chatrchyan:2012xq} and calorimetric jets with $R$~=~0.2 measured by ATLAS~\cite{Aad:2013sla}.}
\end{center}
\end{figure}

In Figure~\ref{fig:v2jetALICE} the measurement of elliptic anisotropy $v_2$ for charged jets with the resolution parameter $R$~=~0.2 is shown in central and semi-central \pbpb collisions. The data show significant positive $v_2$ value in semi-central \pbpb collisions pointing to the path length dependence of jet suppression. In central collisions the current uncertainties on the measurement do not allow to draw a definite conclusion, although the $v_2$ magnitude is also positive. These data are also compared to $v_2$ for full jets measured by ATLAS~\cite{Aad:2013sla} and inclusive charged particles~\cite{Abelev:2012di,Chatrchyan:2012xq}. Although these measurements cannot be directly compared quantitatively due to  different $p_{\rm T}$ scales and centrality selections, qualitatively they agree and provide a clear evidence of path-length dependent parton energy loss.

\subsection{Open heavy flavour and quarkonia production}

\begin{figure}[H]
\begin{center}
\includegraphics[keepaspectratio, width=0.4\columnwidth]{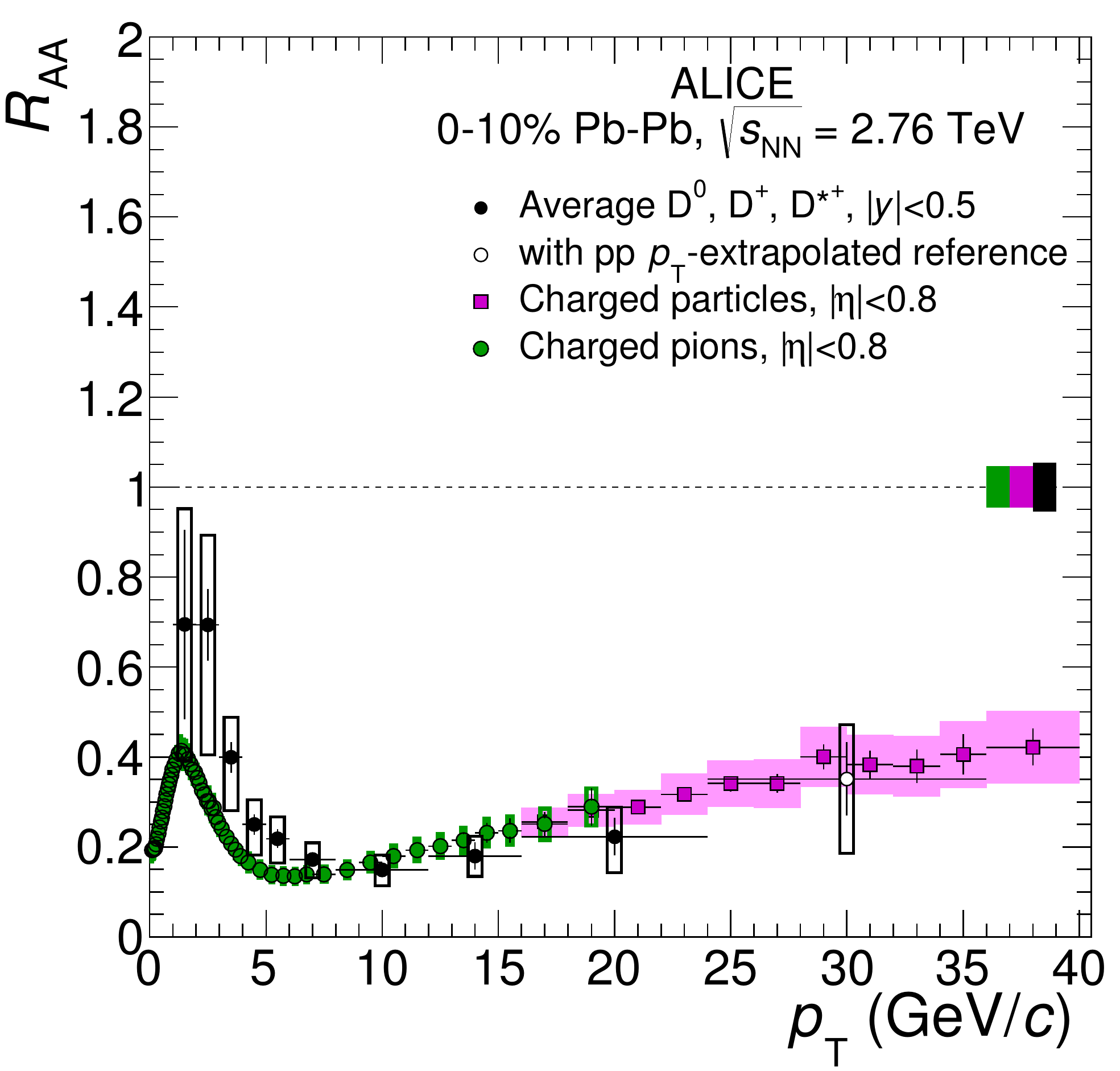} 
\includegraphics[keepaspectratio, width=0.4\columnwidth]{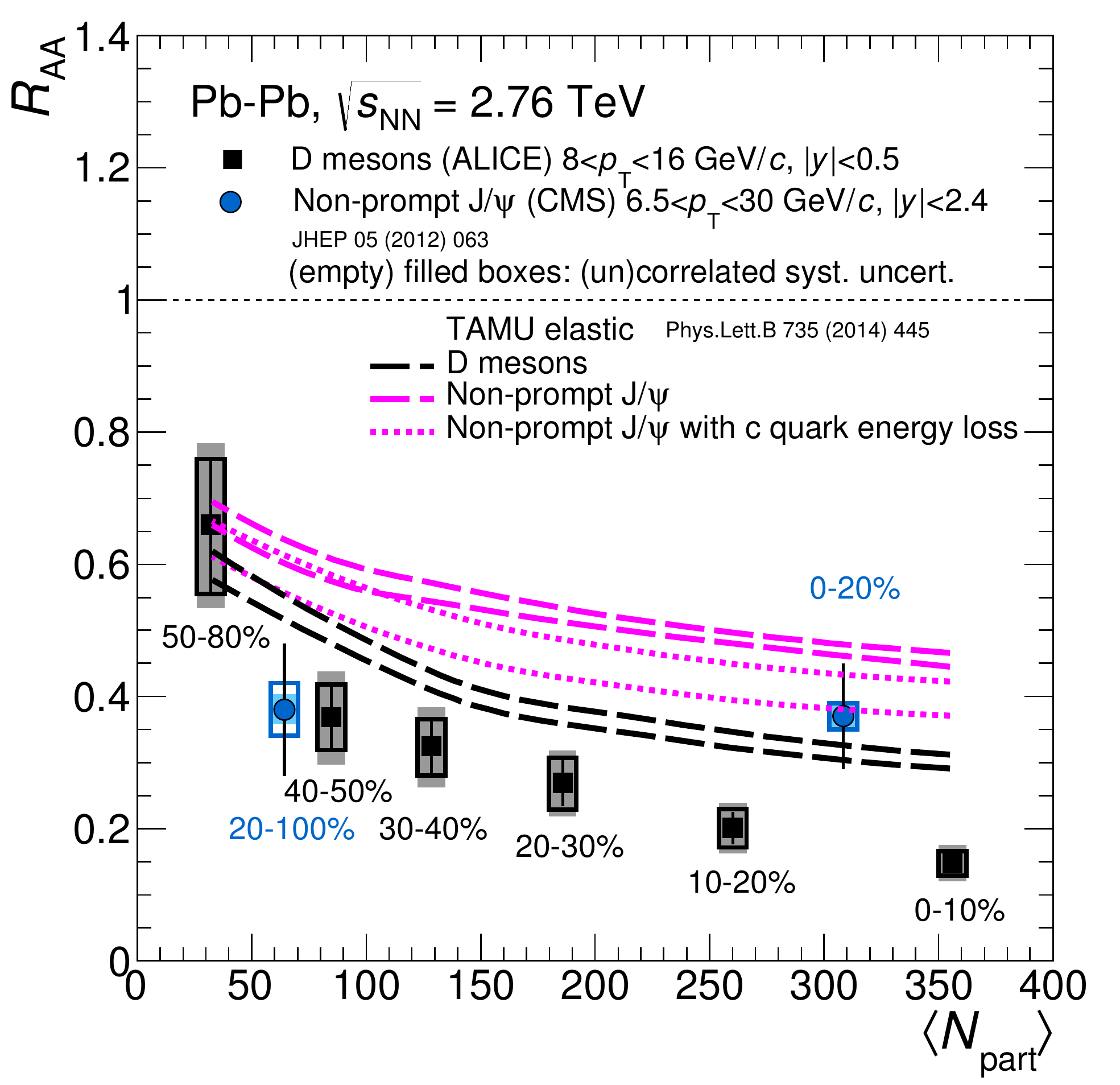} 
\caption{(Color online). D meson nuclear modification factor, $R_{\rm AA}$, in \pbpb
  collisions at $\snnt{2.76}$. Left: $R_{\rm AA}$ as a function
of $p_{\rm T}$ compared to charged hadrons and pions. Right: $R_{\rm AA}$
as a function of $N_{\rm part}$ ~\cite{Adam:2015nna} compared to
non-prompt J/$\psi$ measured by the CMS collaboration~\cite{Chatrchyan:2012np}.}
\label{charged}
\end{center}
\end{figure}

The nuclear modification factors of D mesons~\cite{Adam:2015sza,Adam:2015jda} and electrons from heavy flavour decays~\cite{Abelev:2012qh} at mid-rapidity and of muons from heavy flavour decays at forward rapidity have been measured with ALICE. The results show a strong reduction of the yields (relative to \pp) at large transverse momenta ($p_{\rm T}$~$\geq$~5\,GeV/$c$) in the most central collisions.  Figure~\ref{charged} (left) shows that, within uncertainties and in the measured $p_{\rm T}$ interval, the D meson nuclear modification factor is similar to that of charged pions and
inclusive charged particles. It should be noted that the $R_{\rm AA}$ of D mesons and pions is also sensitive to the shape of the parton momentum distribution and their fragmentation functions. Model calculations including those effects and a colour-charge hierarchy in parton energy loss are able to
describe the measurements~\cite{Djordjevic:2013pba}.

In Fig.~\ref{charged} (right), $R_{\rm AA}^{\rm D}$ as a function of the average number of participant
nucleons~\cite{Adam:2015nna} is shown. This measurement is compared with results from the CMS collaboration of non-prompt J/$\psi$~\cite{Chatrchyan:2012np} and theoretical predictions~\cite{Andronic:2015wma,Nahrgang:2013xaa}. For D mesons, a smaller suppression in peripheral than in central collisions is observed. A larger suppression in central collisions is seen for D mesons than
for non-prompt J/$\psi$, indicating a different energy loss for charm and beauty quarks. This observation is supported by predictions from energy loss models, where the difference
between the $R_{\rm AA}$ of D and B mesons arises from the different masses of $c$ and $b$ quarks. 

\begin{figure}[t!]
\begin{center}
\vspace{-0.5cm}
\includegraphics[keepaspectratio, width=0.8\columnwidth]{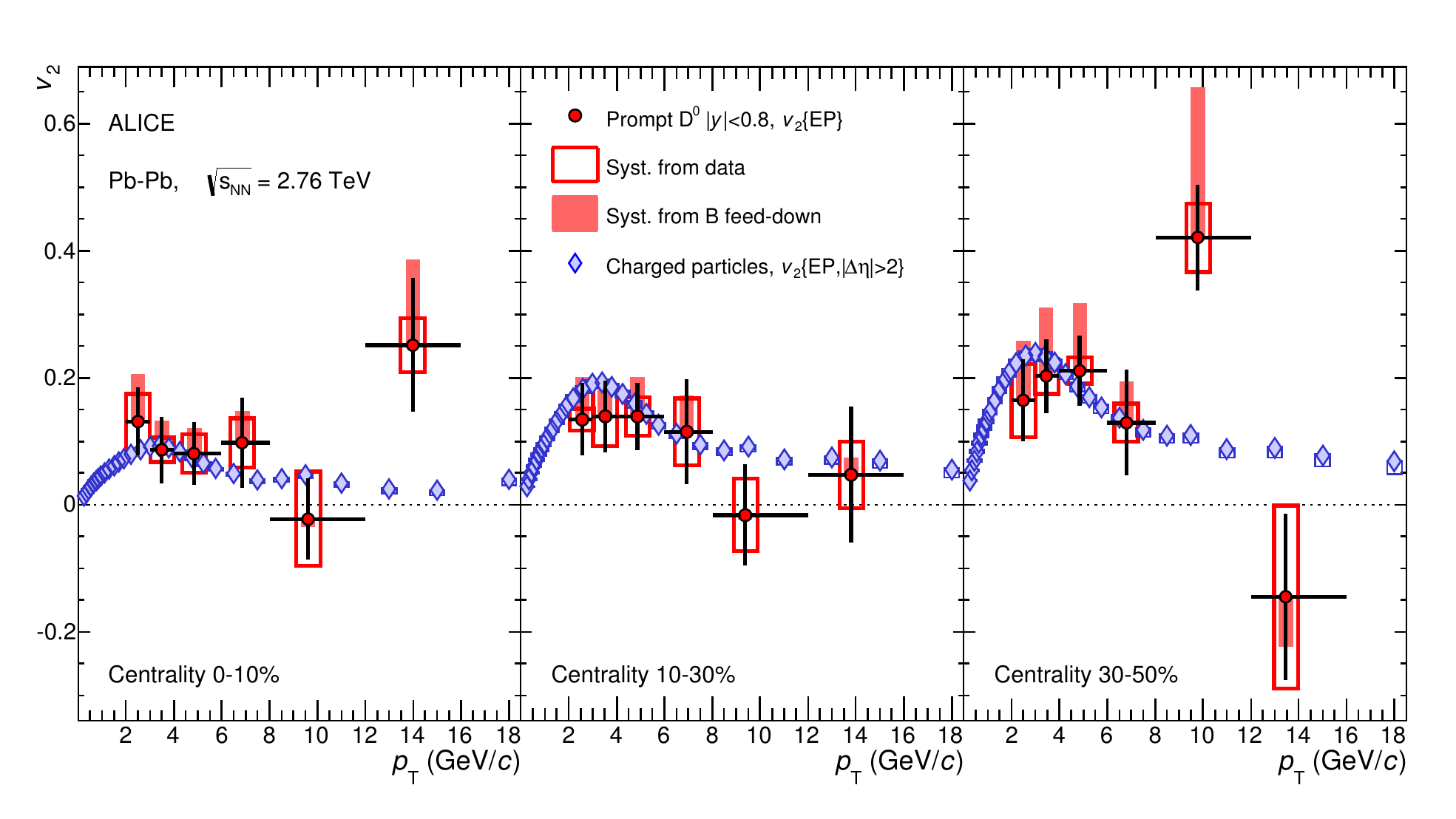}
\caption{(Color online). $\rm D^{0}$ meson $v_{2}$ as a function of $p_{\rm T}$ in
  three centrality ranges and  compared with the $v_{2}$ of charged
  particles~\cite{Abelev:2013lca,Abelev:2014ipa}.}
\label{v2charged}
\end{center}
\end{figure}

The $v_{2}$ of prompt $\rm D^{0}$, $\rm D^{+}$ and $\rm D^{*+}$ mesons at mid-rapidity was measured in three centrality classes: 0-10\%, 10-30\% and 30-50\%~\cite{Abelev:2013lca,Abelev:2014ipa}. The comparisons of the D meson $v_{2}$ as a function of $p_{\rm T}$ with  analogous results for inclusive charged particles are shown in Fig.~\ref{v2charged}. The $v_{2}$ decreases from peripheral to central collisions as expected due to the decreasing initial geometrical anisotropy. The average of the  $v_{2}$ of $\rm D^{0}$, $\rm D^{+}$ and $\rm D^{*+}$ in the centrality class 30-50$\%$ is larger than zero with 5$\sigma$ significance in the range 2~$<$~$p_{\rm T}$~$<$~6\,GeV/$c$. A
positive $v_{2}$ is also observed for $p_{\rm T} >$~6\,GeV/$c$, which most likely originates from the path length dependence of the in-medium partonic energy loss, although the present statistics
does not allow to reach a firm conclusion on this. The measured D meson $v_{2}$ is comparable in magnitude with that of the charged particles, which are mostly light-flavour hadrons. This result indicates that low-$p_{\rm T}$ charm quarks take part in the collective motion of the system. 
 
$R_{\rm AA}$ and $v_{2}$ are two complementary measurements that can be used to better understand the connection between transport coefficients and the description of the in-medium energy loss~\cite{Ayala:2016pvm}.


\subsection{Fluid-like behavior in small systems: \pp and \ppb collisions}
\label{sec:smallsystems}

The study of particle production in high-multiplicity \pp and \ppb collisions at the LHC has revealed unexpected new collective-like phenomena. The understanding of the phenomena is ongoing and different explanations have been proposed~\cite{Gutay:2015cba,Bautista:2015kwa,Dusling:2012iga,Dusling:2012cg,Dusling:2015rja}.

The CMS Collaboration studied the two-particle angular correlations of charged particles in \pp collisions at \sppt{7} and discovered the ridge structure in small systems~\cite{Khachatryan:2010gv}.  While the $p_{\rm T}$-integrated correlation does not show any special feature, in the $p_{\rm T}$ range 1--3\,GeV/$c$ the near side long range angular correlation is clearly observed. Similar structures were also observed in \ppb collisions at \snnt{5.02}~\cite{CMS:2012qk,Abelev:2012ola,Aad:2012gla}. Furthermore, in high multiplicity events, non-zero second-order Fourier coefficients were extracted from the long-range correlations. Using the ALICE capabilities for particle identification,  the proton $v_{2}$ was observed to be smaller than that for pions, up to about $p_{\rm T}=2$\,GeV/$c$~\cite{ABELEV:2013wsa}. This effect is similar to the mass ordering of $v_{2}$ observed in heavy-ion collisions.

The transverse momentum spectra of charged pions, kaons and (anti)protons as a function of the event multiplicity have been measured up to 20\,GeV/$c$~\cite{Adam:2016dau}. At low $p_{\rm T}$ ($<$~$2$--$3$\,GeV/$c$) the spectra exhibit a hardening with increasing multiplicity, this effect is more important for heavy particles than for light particles. We are therefore observing features which resemble the radial flow effects well known from heavy-ion collisions~\cite{Adam:2015kca} and which are well described when a hydrodynamical evolution of the system is considered. At the LHC~\cite{Abelev:2013haa}  it was shown that for high multiplicity \ppb events, the $p_{\rm T}$ spectra were described by the blast-wave function. Using the parameters obtained from the simultaneous fit to pion, kaon, proton and lambda $p_{\rm T}$ spectra the model is able to describe the multi-strange baryon $p_{\rm T}$ distributions ($p_{\rm T}<$~4\,GeV/$c$)~\cite{Adam:2015vsf}. The feature is also observed in \pp collisions simulated with PYTHIA 8~\cite{Corke:2010yf,Ortiz:2013yxa}, where no hydrodynamical evolution is included, instead multiple partonic interactions (MPI) and color reconnection are producing the effects. Some ideas have been proposed to understand the role of MPI in data~\cite{Abelev:2013sqa,Abelev:2014mva,Abelev:2012sk,Cuautle:2015kra,Ortiz:2016kpz}.  The multiplicity dependence of the intermediate to high-$p_{\rm T}$ particle production is studied with the particle ratios~\cite{Adam:2016dau}. The proton-to-pion ratio as a function of the event multiplicity exhibits a maximum (bump) at $p_{\rm T}\approx$~3\,GeV/$c$ and the size of the bump increases with increasing multiplicity. On the other hand, at higher transverse momenta ($p_{\rm T}>$~10\,GeV/$c$) the ratios return to the values measured for \pp and \pbpb collisions. Any particle species dependence of the nuclear modification factor is therefore excluded.

In order to look for the presence of re-scattering effects in high multiplicity \ppb collisions; the K$^{*0}$ and $\phi$ relative to charged kaons production is studied as a function of the cube root of the average mid-rapidity charged particle density.  In heavy-ion collisions the decreasing trend of K$^{*0}$$/$K with increasing fireball size has been explained as a consequence of a re-scattering of K$^{*0}$ decay daughters in the hadronic phase. It is worth noticing that a similar trend is also observed in \ppb collisions~\cite{Adam:2016bpr}.

\begin{figure}[H]
\begin{center}
\includegraphics[width=0.3\textwidth]{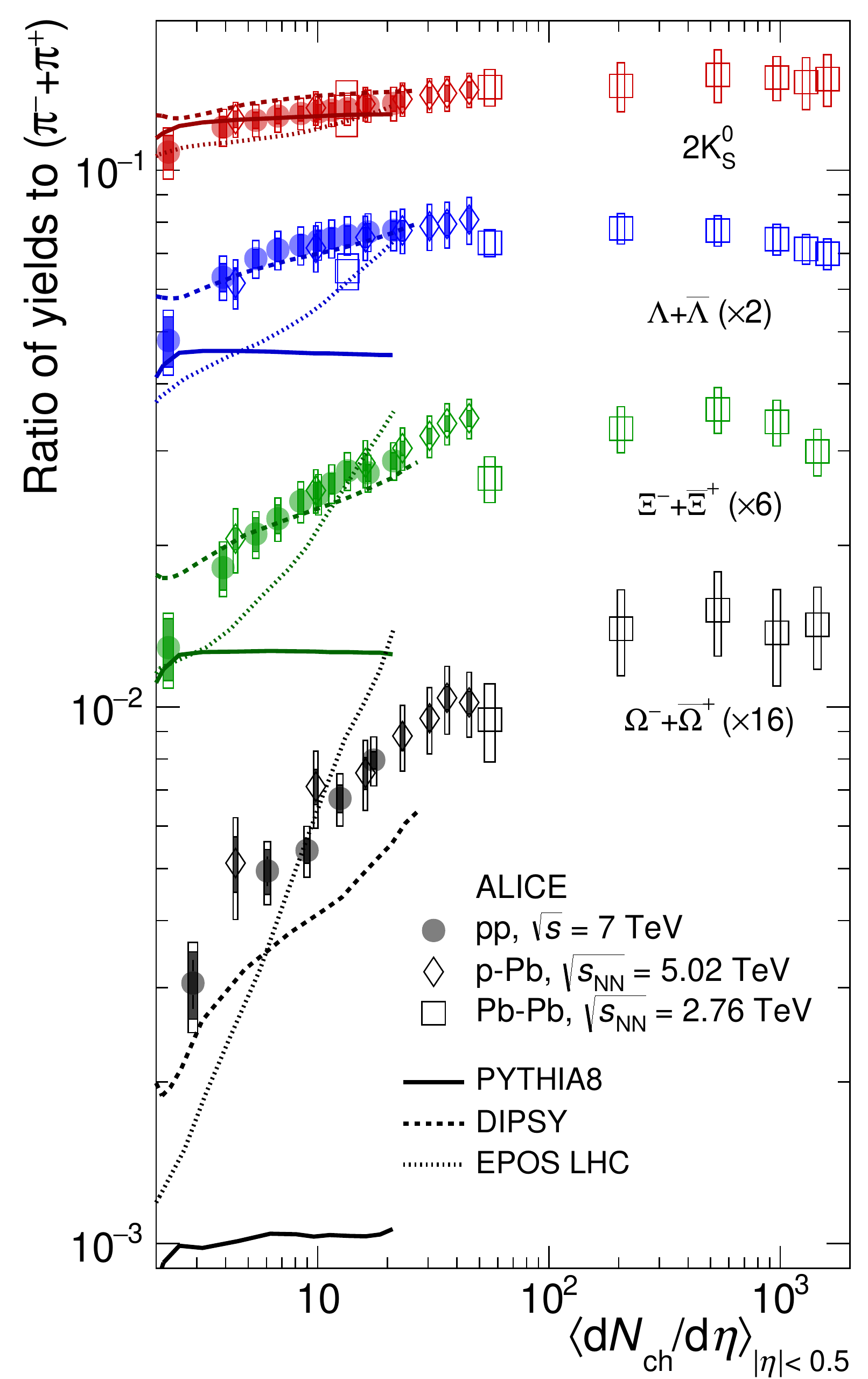}  
\caption{(Color online). Multiplicity dependence of strange  and  multi-strange  hadrons yields normalized to that for charged pions. The empty and dark-shaded boxes show the total systematic uncertainty and the contribution uncorrelated across multiplicity bins, respectively. The values are compared to calculations from MC models and  to  results  obtained  in \pbpb  and  \ppb  collisions at  the  LHC.}
\label{figstr}
\end{center}
\end{figure}

Finally,  we discuss the latest results on multiplicity-dependent enhancement of strange and multi-strange hadron production in \pp and \ppb collisions~\cite{Adam:2016emw}. Figure~\ref{figstr} shows a significant enhancement of strange to non-strange hadron production with increasing particle multiplicity in \pp collisions.   The behaviour observed in \pp collisions resembles that of \ppb collisions at a slightly lower centre-of-mass energy~\cite{Adam:2015vsf}, both in the values of the ratios and in their evolution with the event activity.  This suggests that the origin of strangeness production in hadronic collisions is driven by the characteristics of the event activity rather than by the initial-state collision system or energy. In the context of heavy-ion collisions, this effect (strangeness enhancement) has been considered a signature of the QGP formation. Recently, it has been pointed out that a perfect scaling of the particle ratios with the energy density holds for the different colliding systems~\cite{Cuautle:2016huw} opening new possibilities for a better understanding of the QGP-like features in small systems.

\section{Cosmic ray physics}

Cosmic ray muons are created in Extensive Air Showers (EAS) following the interaction of cosmic ray primaries (protons and heavier nuclei) with nuclei in the upper atmosphere. Primary cosmic rays span a broad  energy  range,  starting  at approximately  $10^{9}$ eV  and  extending  to more  than  $10^{20}$ eV.  The use of high-energy physics detectors for cosmic ray physics was pioneered by ALEPH~\cite{Grupen:2003ih}, DELPHI~\cite{Ridky:2005mx} and L3~\cite{LadrondeGuevara:2004gh} during the Large Electron-Positron (LEP) collider era at CERN. An extension of these earlier studies is now possible at the LHC, where experiments can operate under stable conditions for many years. 

\begin{figure}[h!]
\begin{center}
   \includegraphics[width=0.7\textwidth]{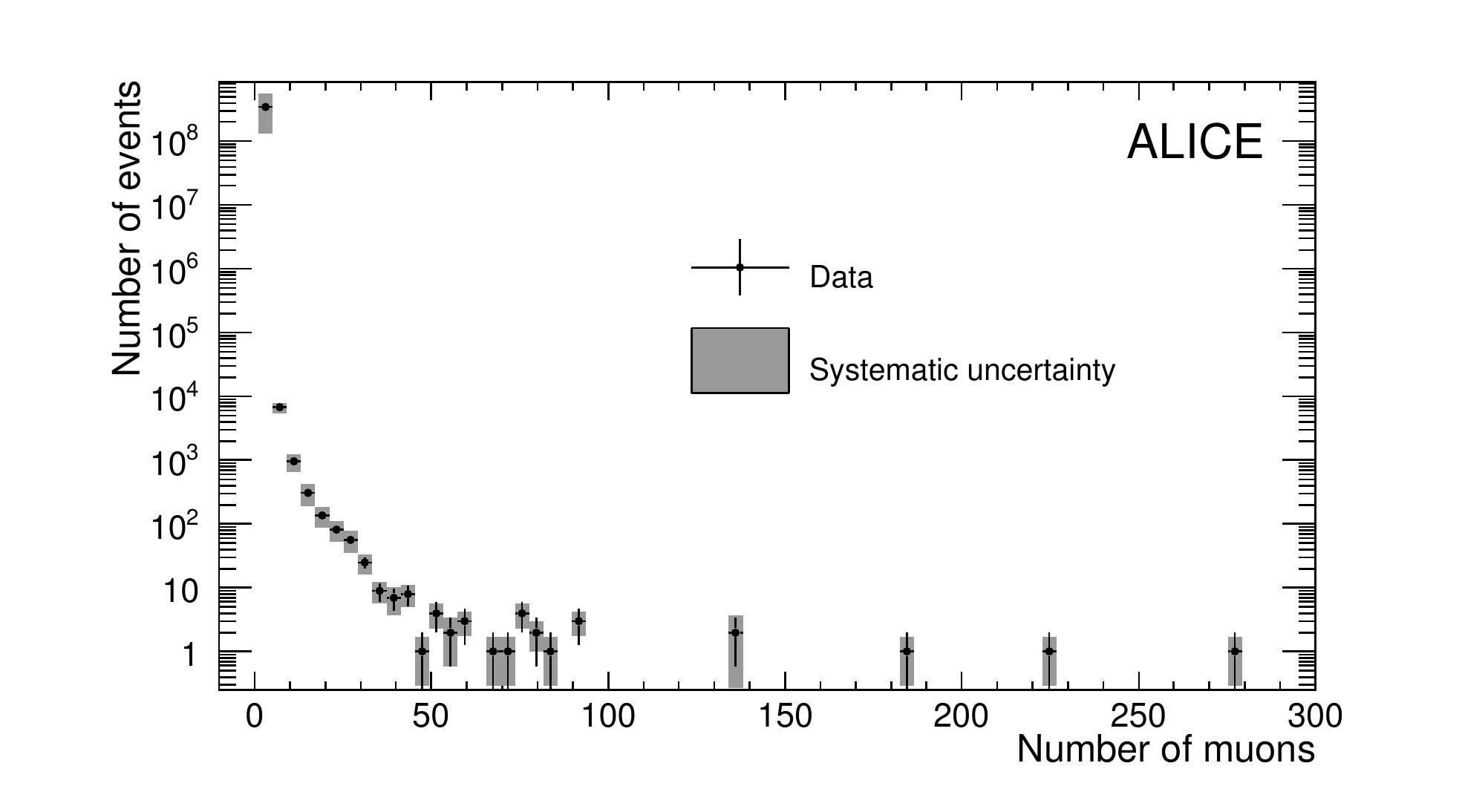} 
   \caption{(Color online). Muon multiplicity distribution of the whole sample of data (2010-2013) corresponding to 30.8 days of data taking.}
  \label{MMD}
\end{center}
\end{figure}

ALICE has also been used to perform studies that are of relevance to astroparticle physics~\cite{ALICE:2015wfa}. Recent studies focus on events containing more than four reconstructed muons in the ALICE TPC, which we refer to as multi-muon events, stem from primaries with energy $E>10^{14}$ eV. 
ALICE dedicated 30.8 days of cosmic ray data  taking recording approximately 22.6 million events containing at least one reconstructed  muon.  
ALICE has reported (see Fig. ~\ref{MMD})  5 events with, $N_{\mu}>100$ and zenith angles less than $50^{o}$  (corresponding to an area density $\rho_{\mu}>5.9$ $m^{ −2}$ in a rate of $1.9\times10^{-6}$ Hz).  

\begin{figure}[h!]
\begin{center}
   \includegraphics[width=0.65\textwidth]{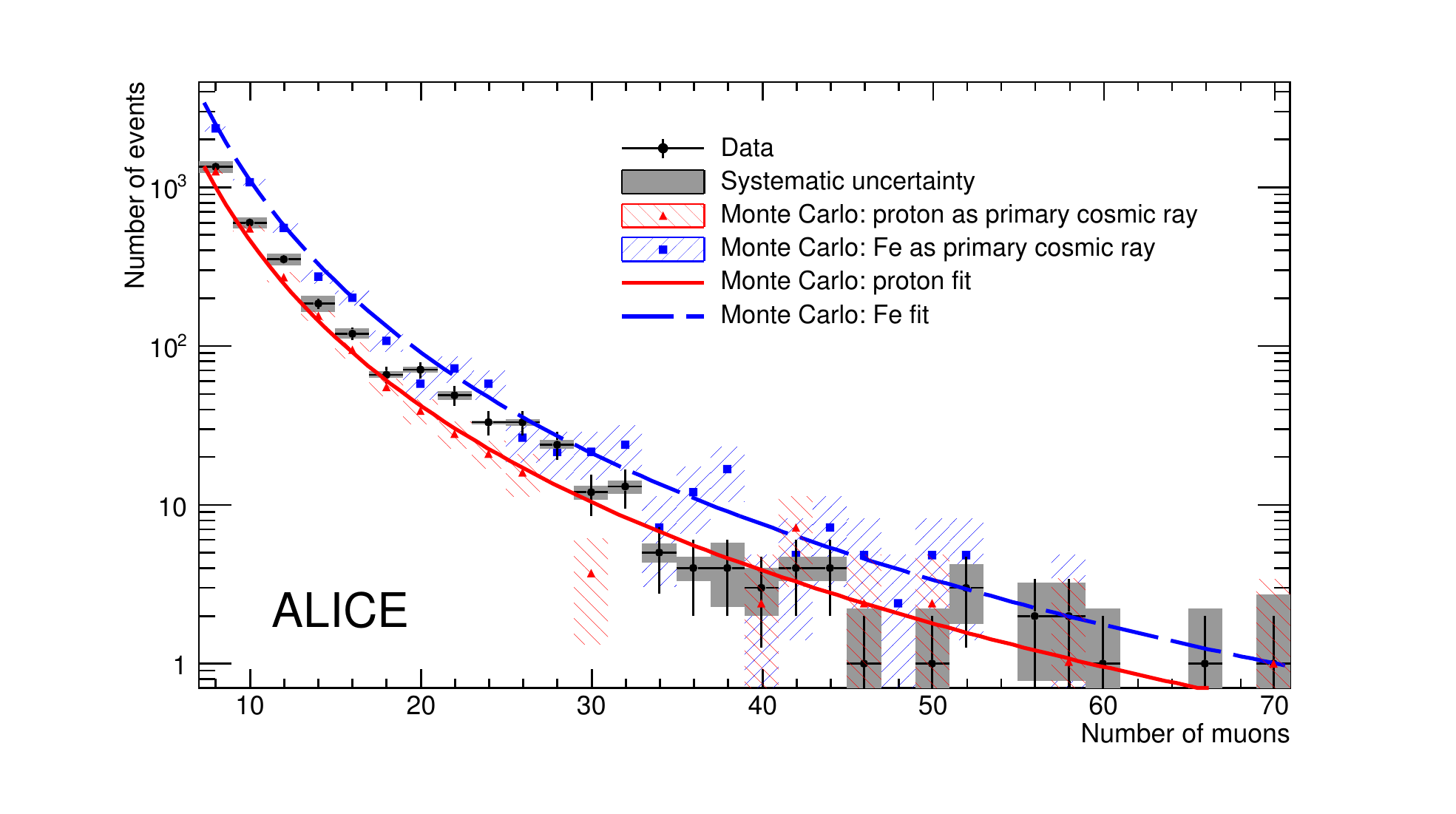}
   \caption{(Color online). The measured MMD compared with the values and fits obtained from CORSIKA simulations with proton and iron primary cosmic rays for 30.8 days of data taking. The errors are shown separately (statistical  and  systematic)  for  data,  while  for  Monte  Carlo  they  are  the  quadrature sum  of  the  statistical  and systematic uncertainties.}
  \label{MMDplusMC}
\end{center}
\end{figure}

Figure ~\ref{MMDplusMC} shows a comparison between data and MC simulations, the result suggests a mixed-ion primary cosmic ray composition with an average mass that increases with energy.

The ALICE measurement  agrees with results from other experiments wich work at the same energy domain (knee).  Following the successful description of the magnitude of the muon multiplicity distribution (MMD) in the low-to-intermediate range of muon multiplicities the same simulation framework was used to study the frequency of high muon multiplicity (HMM) events.

It has been found that the observed rate of HMM events is consistent with the rate predicted by CORSIKA 7350 using QGSJET II-04 to model the development of the resulting airshower, assuming a pure iron composition for the primary cosmic rays.  

Only primary cosmic rays with an energy $E>10^{16}$ eV were found to give rise to HMM events.  The expected rate of HMM events is sensitive to assumptions made about the dominant hadronic production mechanisms in air shower development. The latest version of QGSJET differs from earlier versions in its treatment of forward neutral meson production resulting in a higher muon yield and has been retuned taking into account early LHC results on hadron production  in 7 TeV proton-proton  collisions.
This is the first time that the rate of HMM events, observed at the relatively shallow depth of ALICE,
has been satisfactorily reproduced using a conventional hadronic model for the description of extensive
air showers;  an observation  that places significant  constraints  on alternative,  more exotic  production
mechanisms~\cite{ALICE:2015wfa}.

\section{Outlook}

In this  paper we have presented  the status and some highlights of the ALICE-LHC experiment, including the latest news  on cosmic ray physics and the progress on the characterization of the hot and dense QCD medium created in the heavy-ion collisions. We have discussed several hot topics as the collective-like phenomena in small collision systems. More differential studies are needed in order to reveal the origin of the effects and their impact in our current understanding on the physics of heavy-ion collisions.



\section*{Acknowledgements}

Support  for  this  work  has  been  received  from  CONACYT  under  the  grant  No.   260440 and grant No. 241408;  from DGAPA-UNAM under PAPIIT grant IA102515.

\section*{References}

\bibliographystyle{ws-ijmpe}
\bibliography{biblio_final}

\end{document}